\documentclass{elsart1p}
\usepackage{graphics}
\usepackage{bm}

\begin{document}

\begin{frontmatter} 

\title{$\pi^-$ decay rates of p-shell hypernuclei revisited} 

\author{Avraham Gal}
\ead{avragal@vms.huji.ac.il}
\address{Racah Institute of Physics, The Hebrew University, 
Jerusalem 91904, Israel} 

\date{\today} 

\begin{abstract} 
Explicit expressions for the parity-violating $s$-wave and the 
parity-conserving $p$-wave contributions to $\pi^-$ weak-decay rates of 
$\Lambda$ hypernuclei in the $1p$ shell are given in the weak-coupling limit, 
to update previous shell-model calculations and to compare with recent 
$\pi^-$ spectra and total decay rates measured by the FINUDA Collaboration 
for $_{\Lambda}^7{\rm Li},~_{\Lambda}^9{\rm Be},~_{~\Lambda}^{11}{\rm B}$ 
and $_{~\Lambda}^{15}{\rm N}$. A useful sum rule for the summed strength 
of $\Lambda_{1s}\to p_{1p}$ hypernuclear $\pi^-$ weak decays is derived. 
Fair agreement between experiment and calculations is reached, using the 
primary $s$-wave amplitude and Cohen-Kurath nuclear wavefunctions. The role 
of the $p$-wave amplitude is studied in detail for $_{~\Lambda}^{15}{\rm N}$ 
and found to be secondary. Previous assignments of ground-state spin-parity 
values $J^{\pi}(_{\Lambda}^{7}{\rm Li_{g.s.}})={\frac{1}{2}}^+$ and
$J^{\pi}(_{~\Lambda}^{11}{\rm B_{g.s.}})={\frac{5}{2}}^+$ are confirmed, 
and a new assignment $J^{\pi}(_{~\Lambda}^{15}{\rm N_{g.s.}})={\frac{3}{2}}^+$ 
is made, based on the substantial suppression calculated here for the 
$_{~\Lambda}^{15}{\rm N}({\frac{1}{2}}^+) \to \pi^-~{^{15}{\rm O_{g.s.}}}$ 
weak decay rate. 
\end{abstract}


\end{frontmatter}

\section{Introduction} 
\label{sec:intro}

Mesonic weak decays of $\Lambda$ hypernuclei studied in stopped-$K^-$ 
reactions on nuclear emulsions were used by Dalitz to determine 
ground-state spins of the $s$-shell species 
$_{\Lambda}^3{\rm H}$ and ${_{\Lambda}^4{\rm H}}-{_{\Lambda}^4{\rm He}}$ 
as soon as parity violation had been established in the weak interactions; 
for overview see Refs.~\cite{Gal08,Davis05}. The essence of these early 
calculations was the strong dependence that two-body $\pi^-$ decay branching 
ratios exhibit often on the ground-state (g.s.) spin of the decaying 
hypernucleus. Later, by studying $\pi^-$ angular distributions or 
$\alpha\alpha$ final-state correlations, the ground-state spins of the 
$p$-shell hypernuclei $_{\Lambda}^8{\rm Li}$ and $_{~\Lambda}^{12}{\rm B}$ 
were determined, again from emulsion data \cite{Davis05}. The advent of 
counter experiments using the ($K^-,\pi^-$) and ($\pi^+,K^+$) reactions 
allows a systematic study of weak decays for other species not readily or 
uniquely accessible in emulsion work. Thus, the $\pi^-$ weak decay rates 
measured for the $p$-shell hypernuclei $_{~\Lambda}^{11}{\rm B}$ and 
$_{~\Lambda}^{12}{\rm C}$, both produced on $^{12}$C \cite{Sato05}, 
suggest on comparing to the calculations by Motoba et al. \cite{Motoba88,
Motoba92,Motoba94} that $J^{\pi}(_{~\Lambda}^{11}{\rm B_{g.s.}})=
{\frac{5}{2}}^+$ and $J^{\pi}(_{~\Lambda}^{12}{\rm C_{g.s.}})=1^-$. Recently, 
the spin-parity value 
$J^{\pi}(_{\Lambda}^7{\rm Li}_{\rm g.s.})={\frac{1}{2}}^+$ was determined 
at KEK in a $^7{\rm Li}(\pi^+,K^+\gamma)$ experiment \cite{Sasao04}, 
comparing the yield of $\gamma$ rays subsequent to the weak decay 
$_{\Lambda}^7{\rm Li}_{\rm g.s.}\to\pi^-~{^7{\rm Be}^{\ast}(429~{\rm keV})}$ 
again with the calculations of Motoba et al. Ground-state spin values 
of light $\Lambda$ hypernuclei provide valuable information on the spin 
dependence of the $\Lambda N$ interaction and on $\Lambda N-\Sigma N$ 
coupling effects in hypernuclei \cite{Mill07}. For a recent review and 
compilation of hypernuclear data, see Ref.~\cite{Hash06}. 

Very recently, at PANIC08, $\pi^-$ spectra of weakly decaying $\Lambda$ 
hypernuclei have been presented by the FINUDA Collaboration for 
$_{\Lambda}^7{\rm Li},~ _{\Lambda}^9{\rm Be},~_{~\Lambda}^{11}{\rm B},~ 
_{~\Lambda}^{15}{\rm N}$ \cite{Botta09}. 
Whereas individual final nuclear states cannot be resolved in these 
decay spectra, the $_{~\Lambda}^{11}{\rm B}$ spectrum shows evidence 
for two groups of states separated by about 6 MeV in the residual 
$^{11}{\rm C}$ nucleus, and the $_{~\Lambda}^{15}{\rm N}$ spectrum 
is clearly dominated by $^{15}{\rm O}_{\rm {g.s.}}$. 
Besides these features that concern the shape of the $\pi^-$ spectrum, 
the most meaningful entity to compare at present between experiment 
and theory is the total rate $\Gamma_{\pi^-}^{\rm tot}({_{\Lambda}^A}Z)$
for $\pi^-$ decay from the ground state of ${_{\Lambda}^A}Z$. In hypernuclei, 
owing to Pauli blocking for the low-momentum recoil proton (with $q \sim q_0 
= 101$ MeV/c, where the momentum $q_0$ holds for $\Lambda \to \pi^- p$), 
the $\pi^-$ decay rate $\Gamma_{\pi^-}^{\rm tot}({_{\Lambda}^A}Z)$ drops 
steadily with $A$ away from the corresponding free-space decay rate 
$\Gamma_{\pi^-}^{\rm free}$.{\footnote{$\Gamma_{\pi^-}^{\rm free}/
\Gamma_{\Lambda}=0.639$ in terms of the $\Lambda$ decay width 
$\Gamma_{\Lambda}=\hbar/\tau_{\Lambda}$, where 
$\tau_{\Lambda}=(2.631\pm 0.020)\times 10^{-10}$~s for the $\Lambda$ 
mean life \cite{PDG08}.}} The FINUDA data indicate a fall-off of 
$\Gamma_{\pi^-}^{\rm tot}({_{\Lambda}^A}Z)/\Gamma_{\Lambda}$ 
by a factor of three throughout the $1p$ shell, from about 0.35 for 
$_{\Lambda}^7{\rm Li}$ down to about 0.11 for $_{~\Lambda}^{15}{\rm N}$. 
The data bear statistical uncertainties between approximately $15\%$ 
and $30\%$ and, except for $_{~\Lambda}^{11}{\rm B}$, considerably 
smaller systematic uncertainties. 

Pionic decays of light hypernuclei have been studied particularly for 
$_{\Lambda}^5{\rm He}$ and $_{~\Lambda}^{12}{\rm C}$, focusing on the 
renormalization of the $\Lambda \to N \pi$ decay vertex in the nuclear 
medium; for comprehensive reviews, see Refs.~\cite{OR98,AG02}. 
The present work, however, is related closely to spectroscopic aspects 
of pionic decays as studied systematically by Motoba et al., with application 
in Ref.~\cite{Motoba88} particularly to decays of $p$-shell hypernuclei, 
and with update and extension in Refs.~\cite{Motoba92,Motoba94}. 
In these calculations, the pion final-state interaction was incorporated 
by using pion-nuclear distorted waves, and the structure of the nuclear 
core was treated by using the Cohen-Kurath (CK) spectroscopic calculations 
\cite{CK65,CK67}. For $p$-shell hypernuclei, it was found that the total 
$\pi^-$ decay rate is dominated by transitions $\Lambda_{1s}\to p_{1p}$, 
with little strength left for transitions to higher nuclear configurations 
dominated by $\Lambda_{1s}\to p_{2s,1d}$ transitions. 
Here we follow this approach, providing {\it explicit} expressions for 
$\Lambda_{1s}\to p_{1p}$ transitions from $p$-shell hypernuclear ground states 
in the weak-coupling limit to final nuclear states within the $1p$ shell. 
The nuclear states are given by the CK wavefunctions, specifically in terms of 
the published CK coefficients of fractional parentage \cite{CK67}. We discuss 
the choice of nuclear form factors involved in these transitions, with 
a parametrization that fully accounts for the distortion of the outgoing pion 
as calculated by Motoba et al. \cite{Motoba88,Motoba92,Motoba94}. Furthermore, 
we derive and demonstrate the use of a new sum rule which encapsulates the 
suppressive effect of the Pauli principle on the total $\pi^-$ weak decay 
rate. Our primary aim is to allow any concerned experimenter to check on 
his/her own the calculational state of the art in $\pi^-$ weak decays of light 
hypernuclei. 

In the calculations reported in the present work, we have checked that for the 
main $\Lambda_{1s} \to p_{1p}$ transitions, as well as for the summed 
strength, it suffices to consider the leading $s$-wave amplitude. The $p$-wave 
amplitude produces an observable effect only for decays that are suppressed 
within the $s$-wave approximation, as is demonstrated here in detail for the 
decay of $_{~\Lambda}^{15}{\rm N}({\frac{1}{2}}^+)$. The neglect of the 
$p$-wave weak decay amplitude incurs errors of less than $10\%$ when the main 
transitions are summed upon, consistently with the stated precision of our 
calculations. A comparison is then made between the calculated $\pi^-$ decay 
rates and the rates measured by FINUDA, based on which the hypernuclear g.s. 
spin-parity assignments 
$J^{\pi}(_{\Lambda}^{7}{\rm Li_{g.s.}})={\frac{1}{2}}^+$ and 
$J^{\pi}(_{~\Lambda}^{11}{\rm B_{g.s.}})={\frac{5}{2}}^+$ are confirmed, and 
a new assignment $J^{\pi}(_{~\Lambda}^{15}{\rm N_{g.s.}})={\frac{3}{2}}^+$ 
is established.

\section{Methodology and Input} 

\subsection{$\Lambda \to N \pi$ decay input} 

The free $\Lambda$ weak decay rate is dominated to $99.6\%$ by the 
nonleptonic, mesonic decays $\Lambda \to \pi^- p, \pi^0 n$: 
\begin{equation} 
\Gamma_{\Lambda} \approx \Gamma_{\pi^-}^{\rm free} + \Gamma_{\pi^0}^{\rm free}
~~~~~(\Gamma_{\pi^-}^{\rm free} : \Gamma_{\pi^0}^{\rm free} \approx 2 : 1),   
\label{eq:GammaLambda} 
\end{equation} 
where each one of these partial rates consists of parity-violating $s$-wave 
and parity-conserving $p$-wave terms: 
\begin{equation} 
\Gamma_{\pi^{\alpha}}^{\rm free}=c_\alpha\frac{q_0}{1+\omega_{\pi}(q_0)
/E_N(q_0)}(|s_{\pi}|^2+|p_{\pi}|^2\frac{q^2}{q_0^2}),~~~~~ 
|\frac{p_{\pi}}{s_{\pi}}|^2 \approx 0.132, 
\label{eq:Gammapi} 
\end{equation} 
with $\omega_{\pi}(q_0)$ and $E_N(q_0)$ the energies of the free-space 
decay pion and the recoil nucleon, respectively. The ratio $2:1$ in 
Eq.~(\ref{eq:GammaLambda}), $c_{-}/{c_0}\approx 2$ in terms of the strength 
parameters $c_\alpha$ in Eq.~(\ref{eq:Gammapi}), is a consequence of the 
$\Delta I=1/2$ rule, satisfied empirically by the weak interactions. 

\subsection{$\pi^-$ decay rates for $\Lambda_{1s}\to p_{1p}$ transitions} 

We consider $\pi^-$ decay from an initial hypernuclear state 
$|i> \equiv |{_{\Lambda}^A}Z;\alpha_i,T_i \tau_i,J_iM_i>$ to a final nuclear 
state $|f> \equiv |{^A}(Z+1);\alpha_f,T_f \tau_f,J_fM_f>$ in the nuclear 
$1p$ shell, where $J$ ($T$) and $M$ ($\tau$) stand for the total angular 
momentum (isospin) and their z projections, and $\alpha$ stands for any 
other quantum numbers providing spectroscopic labels. The $\pi^-$ decay 
rate for a $\Lambda_{1s}\to p_{1p}$ transition induced by the $s$-wave 
amplitude is given by 
\begin{equation} 
\Gamma^{(\rm s)}_{\pi^-}(i \to f)=c_{-}\frac{q}{1+\omega_{\pi^-}(q)/E_A(q)}
|s_{\pi}|^2~{\cal P}^{(\rm s)}_{i \to f}, 
\label{eq:Gammapif} 
\end{equation} 
where the effective proton number ${\cal P}$ for this transition 
is defined by 
\begin{equation} 
{\cal P}^{(\rm s)}_{i \to f}=\int{\frac{d\Omega_{\bf q}}{4\pi}}
{\frac{1}{(2J_i+1)}}
\sum_{M_iM_f}|<f|\int{d^3{\bf r}\chi_{\bf q}^{(-)\ast}(\bf r)} 
\sum_{k=1}^{A}{\delta({\bf r}-{\bf r_k})V_{k+}}|i>|^2.  
\label{eq:calP} 
\end{equation} 
In Eq.~(\ref{eq:calP}), $\chi_{\bf q}^{(-)}$ is an incoming pion distorted 
wave (DW), $\exp(-{\rm i}{\bf q}\cdot{\bf r})$ in the plane-wave (PW) limit, 
and the $V$-spin raising operator $V_{+}$ transforms $\Lambda$s to protons: 
$V_{+}|\Lambda>=|p>$. The summation on $k$ extends over the nucleons of the 
nuclear decay product. Eq.~(\ref{eq:calP}) reduces to 
${\cal P}^{(\rm s)}_{i\to f}={\cal S}^{(\rm s)}_{i\to f}|F_{\rm DW}^{(\rm s)}
(q)|^2$, where the spectroscopic factor for the transition $\Lambda_{1s}(i)\to 
p_{1p}(f)$ associated with the $\pi^-$ decay $s$-wave amplitude is given by   
\begin{eqnarray} 
{\cal S}^{(\rm s)}_{i \to f} & = & N_f^{(1p)}~\langle T_i\tau_i,\frac{1}{2} 
\frac{1}{2}; T_f \tau_f {\rangle}^2~(2J_f+1)~ (~ \sum_j ~[\alpha_i,T_i,J_c; 
\frac{1}{2},j|\}\alpha_f,T_f,J_f] \nonumber \\ 
& & \times (-1)^{J_c+j+J_f} \sqrt{(2j+1)} 
\left\{\begin{array}{ccc}j&J_f&J_c \\ J_i&\frac{1}{2}&1 \\ 
\end{array}\right\}~ )^2. 
\label{eq:calPjj} 
\end{eqnarray} 
Here $J_c$ is the total angular momentum of the initial core nucleus which 
couples with the $s$-shell $\Lambda$ spin 1/2 to yield $J_i$ and with the 
$p$-shell proton angular momentum $j=1/2,~3/2$ to yield $J_f$. 
In Eq.~(\ref{eq:calPjj}), $N_f^{(1p)}$ which is the number of $p$-shell 
{\it nucleons} in the final state is followed by a squared isospin 
Clebsch-Gordan (CG) coefficient accounting for $\Lambda \to p$ transitions; 
the first symbol in the sum over $j$ is a fractional-parentage coefficient 
(CFP), and the curly bracket stands for a $6j$ symbol. The phase factor 
$(-1)^{J_c+j+J_f}$ is consistent with the CK CFPs. Eq.~(\ref{eq:calPjj}) 
corrects the corresponding expression Eq.~(8a) used in Ref.~\cite{ZD75} 
for the special case of $_{~\Lambda}^{11}{\rm B}$. We note that it is 
straightforward to transform ${\cal S}^{(\rm s)}_{i\to f}$, 
Eq.~(\ref{eq:calPjj}), from a $jj$-coupling representation to $LS$,
a representation that is quite useful in the beginning of the $1p$ shell.

The DW form factor $F_{\rm DW}^{(\rm s)}$ is defined by 
\begin{equation} 
F_{\rm DW}^{(\rm s)}(q)=\int_{0}^{\infty}{u_{1p}^N(r){\tilde j}_1(qr)
u_{1s}^{\Lambda}(r)r^2~dr}, 
\label{eq:DWFF}  
\end{equation} 
where $u_{1p}^N$ and $u_{1s}^{\Lambda}$ are the radial wavefunctions of the 
$1p$ nucleon and the $1s$ $\Lambda$, respectively, and the pion radial DW 
${\tilde j}_1(qr)$ goes over to the spherical Bessel function $j_1(qr)$ in 
the plane-wave (PW) limit. Our definition of form factor, Eq.~(\ref{eq:DWFF}), 
is related to the squared form factor $\eta_s(1p_j;q)$ in Eq.~(4.6) of 
Ref.~\cite{Motoba88} by $|F_{\rm DW}^{(\rm s)}(q)|^2=\eta_s(1p_j;q)$, 
where the dependence on $j$, for a given $l=1$, is here suppressed. 
Rather, a state dependence of the form factor arises from the implicit 
dependence of $F_{\rm DW}^{(\rm s)}(q)$ on the $\Lambda$ and proton binding 
energies which vary along the $p$ shell, and will be considered in the next 
subsection. 

\subsubsection{Digression to $p$-wave amplitudes} 

For completeness, as well as for use in the case of $_{~\Lambda}^{15}{\rm N}$ 
below, we record relevant expressions for $p$-wave amplitudes in analogy 
with the $s$-wave expressions listed above. Thus, in addition to 
$\Gamma^{(\rm s)}_{\pi^-}(i \to f)$ of Eq.~(\ref{eq:Gammapif}) we have 
\begin{equation} 
\Gamma^{(\rm p)}_{\pi^-}(i \to f)=c_{-}\frac{q}{1+\omega_{\pi^-}(q)/
E_A(q)}~|p_{\pi}|^2~\frac{q^2}{q_0^2}~{\cal P}^{(\rm p)}_{i \to f}, 
\end{equation} 
where ${\cal P}^{(\rm p)}_{i \to f}={\cal S}^{(\rm p)}_{i\to f}
|F_{\rm DW}^{(\rm p)}(q)|^2$, and the spectroscopic factor for the transition 
$i \to f$ associated with the $\pi^-$ decay $p$-wave amplitude is given by 
\begin{eqnarray} 
{\cal S}^{(\rm p)}_{i \to f} & = & N_f^{(1p)}~\langle T_i\tau_i,\frac{1}{2} 
\frac{1}{2}; T_f \tau_f {\rangle}^2~(\delta_{J_iJ_f}~[\alpha_i,T_i,J_c;
\frac{1}{2},j=\frac{1}{2}|\}\alpha_f,T_f,J_f]^2 \nonumber \\ & & + 
4~(2J_f+1)\left\{\begin{array}{ccc}2&\frac{1}{2}&\frac{3}{2} \\ J_c&J_f&J_i \\ 
\end{array}\right\}^2 [\alpha_i,T_i,J_c;\frac{1}{2},j=\frac{3}{2}|\}
\alpha_f,T_f,J_f]^2). 
\label{eq:calPjjpwave}
\end{eqnarray} 
We note the approximation adopted here of using a single $p$-wave form 
factor $F_{\rm DW}^{(\rm p)}(q)$, following Ref.~\cite{Motoba88} which 
in the relevant range of $q$ values, furthermore, suggests that 
$|F_{\rm DW}^{(\rm p)}(q)|^2\approx |F_{\rm DW}^{(\rm s)}(q)|^2$ to better 
than $10\%$.

\subsection{$\Lambda_{1s}\to p_{1p}$ transition form factor} 

The DW transition form factor (\ref{eq:DWFF}) is evaluated as follows. 
First, we compute the PW form factor using harmonic-oscillator shell-model 
wavefunctions, with appropriate parameters $\nu = 0.41$ fm$^{-2}$ 
for a $p$-shell nucleon and $\lambda = 0.33$ fm$^{-2}$ for a $s$-shell 
$\Lambda$ hyperon \cite{DG78}: 
\begin{equation} 
|F_{\rm PW}^{(\rm s)}(q)|^2=\frac{q^2}{6\lambda}~\left 
(\frac{2\sqrt{\nu\lambda}}{\nu+\lambda}\right )^5 \exp\left 
(-\frac{q^2}{\nu+\lambda}\right ), 
\label{eq:PWFF} 
\end{equation} 
yielding $|F_{\rm PW}^{(\rm s)}(q_0)|^2=0.0898$ for the free-space 
$\Lambda \to \pi^- p$ decay center-of-mass momentum $q=q_0=100.7$ MeV/c. 
A more refined evaluation using 
a two-term Gaussian $\Lambda$ wavefunction fitted in Ref.~\cite{GSD71} to 
$\Lambda$ binding energies in the $1p$ shell gives a value 
$|F_{\rm PW}^{(\rm s)}(q_0)|^2=0.0876$ which we will adopt here, compared with 
$|F_{\rm PW}^{(\rm s)}(q_0)|^2 \approx 0.085$ for more realistic wavefunctions 
as estimated from Fig.~2 of Ref.~\cite{Motoba88}. We then estimate the ratio 
$|F_{\rm DW}^{(\rm s)}(q)|^2/|F_{\rm PW}^{(\rm s)}(q)|^2$ by comparing PW 
rates from Table 6, Ref.~\cite{Motoba88}, with the corresponding DW rates 
from Table 1, Ref.~\cite{Motoba94} (FULL V.R. entry), providing an average 
value $1.47 \pm 0.16$ for this ratio. Since both $|F_{\rm PW}^{(\rm s)}(q)|^2$ 
and $|F_{\rm DW}^{(\rm s)}(q)|^2$ rise approximately linearly with $q$ around 
$q_0$, with similar slopes, the deviations from the average ratio reflect 
primarily the variation of $\Lambda$ and proton wavefunctions across the 
$1p$ shell, from $_{\Lambda}^7{\rm Li}$ to $_{~\Lambda}^{15}{\rm N}$. 
This uncertainty, of order $10\%$, is likely to be the largest one in the 
present calculation of $\pi^-$ decay rates. In the present update we employ 
the following average value for $|F_{\rm DW}^{(\rm s)}(q)|^2$: 
\begin{equation} 
|F_{\rm DW}^{(\rm s)}(q)|^2 \approx 1.47 \times |F_{\rm PW}^{(\rm s)}(q)|^2~ 
\approx 0.129 \times (1~+~1.296 \times{\frac{q-q_0}{q_0}}), 
\label{eq:DWFFinal} 
\end{equation} 
where the linear $q$ dependence follows by expanding Eq.~(\ref{eq:PWFF}) for 
$|F_{\rm PW}^{(\rm s)}(q)|^2$ about $q_0$.

\subsection{Sum rules for $\Lambda_{1s}\to p_{1p}$ transitions} 

Defining $s$-wave and $p$-wave $(2J_i+1)$-averaged spectroscopic strengths 
and $(2J_i+1)$-averaged effective proton numbers in the nuclear core of the 
decaying hypernuclear ground state, 
\begin{equation} 
{\cal S}_{{\bar i} \to f} \equiv 
\sum_{J_i}\frac{(2J_i+1)}{2(2J_c+1)}{\cal S}_{i \to f},~~~
{\cal P}_{{\bar i} \to f} \equiv 
\sum_{J_i}\frac{(2J_i+1)}{2(2J_c+1)}{\cal P}_{i \to f},
\label{eq:av}  
\end{equation} 
it becomes possible to derive a sum rule involving all final 
states within the $1p$ shell. The $(2J_i+1)$-average in Eq.~(\ref{eq:av}) 
is over the two initial $J_i=J_c \pm 1/2$ values for a given $J_c$ (except 
for $J_c=0$ which implies a unique $J_i=1/2$). 
For $T_i=T_c=0$ which holds for the present application, 
$T_f=1/2$ and the CG coefficient in Eqs.~(\ref{eq:calPjj}) and 
(\ref{eq:calPjjpwave}) assumes the value 1. We demonstrate the sum-rule 
derivation for transitions associated with the $\pi^-$ decay $s$-wave 
amplitude. Noting that by orthogonality, 
\begin{equation} 
\sum_{J_i}(2J_i+1)\left\{\begin{array}{ccc}j&J_f&J_c \\ J_i&\frac{1}{2}&1 \\ 
\end{array}\right\}\left\{\begin{array}{ccc}j'&J_f&J_c \\ J_i&\frac{1}{2}&1 \\
\end{array}\right\}=\frac{1}{(2j+1)}\delta_{jj'},  
\label{eq:6j}  
\end{equation} 
a $(2J_i+1)$-average application to the right-hand side of 
Eq.~(\ref{eq:calPjj}) yields: 
\begin{equation} 
{\cal S}^{(\rm s)}_{{\bar i} \to f}=
\frac{1}{4}\sum_{j}\frac{(2J_f+1)(2T_f+1)}{(2J_c+1)(2T_c+1)}~N_f^{(1p)}~
[\alpha_i,T_i,J_c;\frac{1}{2},j|\}\alpha_f,T_f,J_f]^2, 
\label{eq:averageJi} 
\end{equation} 
where we used implicitly the isospin values $T_i=T_c=0,~T_f=1/2$. 
We then sum over $\alpha_f,T_f,J_f$ using a well known 
stripping sum rule (Ref.~\cite{CK67}, Eq.~(6)): 
\begin{eqnarray} 
\sum_{\alpha_f,T_f,J_f}&\frac{(2J_f+1)(2T_f+1)}{(2J_c+1)(2T_c+1)}&N_f^{(1p)}
[\alpha_i,T_i,J_c;\frac{1}{2},j|\}\alpha_f,T_f,J_f]^2 \nonumber \\ 
 = & 2(2j+1)-N_{c(j)}^{(1p)}, 
\label{eq:stripping} 
\end{eqnarray} 
where $N_{c(j)}^{(1p)}$ is the number of $1p_j$ nucleons in the initial 
(nuclear core) state. Summing over $j=l\pm 1/2$ as indicated on the right-hand 
side of Eq.~(\ref{eq:averageJi}), we obtain the sum rules: 
\begin{equation} 
\sum_{\alpha_f,T_f,J_f}{\cal S}^{(\rm s)}_{{\bar i} \to f} = 
3(1-\frac{1}{12}N_c^{(1p)}),~~
\sum_{\alpha_f,T_f,J_f}{\cal P}^{(\rm s)}_{{\bar i} \to f} \approx 
(1-\frac{1}{12}N_c^{(1p)})~3|F_{\rm DW}^{(\rm s)}(\bar q)|^2,  
\label{eq:sumrule} 
\end{equation} 
where $N_c^{(1p)}=\sum_j{N_{c(j)}^{(1p)}}$ is the number of $p$-shell nucleons 
in the nuclear core state, and $\bar q$ is an appropriate value of the 
decaying pion momentum conforming with the closure approximation implied in 
deriving the sum rule for $\cal P$. The right-hand sides in 
Eqs.~(\ref{eq:sumrule}) display the suppressive effect of the Pauli principle 
through the factor $(1-N_c^{(1p)}/12)$ which decreases from 1 (no suppression) 
for the decay of $_{\Lambda}^5{\rm He}$ into the fully available $1p$ proton 
shell ($N_c^{(1p)}=0$) to 0 (full suppression) for the decay of 
$_{~\Lambda}^{16}{\rm O}$ into the fully occupied $1p$ proton shell. 
And finally, the factor 3 in front of $|F_{\rm DW}^{(\rm s)}|^2$ stands for 
the $2l+1=3$ $m$-states available to the final $p$-shell nucleon. 

By following similar procedures, it can be shown that precisely the same 
structure of sum rule, Eq.~(\ref{eq:sumrule}), holds also for transitions 
associated with the $\pi^-$ decay $p$-wave amplitude: 
\begin{equation} 
\sum_{\alpha_f,T_f,J_f}{\cal P}^{(\rm p)}_{{\bar i} \to f} \approx 
(1-\frac{1}{12}N_c^{(1p)})~3|F_{\rm DW}^{(\rm p)}(\bar q)|^2.  
\label{eq:sumrulepwave} 
\end{equation}

\subsection{Summary of formulae} 

For comparison with measured $\pi^-$ decay rates, our $s$-wave results are 
summarized below, assuming that the effect of the neglected 
$p$-wave largely cancels out, particularly for {\it summed} rates. 
Thus, the decay rate for a specific transition $i \to f$ is given by 
\begin{equation} 
\frac{\Gamma_{\pi^-}(i \to f)}{\Gamma_{\Lambda}} \approx 0.639~\frac{q}{q_0}~ 
\frac{1+\omega_{\pi}(q_0)/E_N(q_0)}{1+\omega_{\pi^-}(q)/E_A(q)}~
{\cal P}^{(\rm s)}_{i \to f} \approx 0.743~\frac{q}{q_0}~
{\cal P}^{(\rm s)}_{i \to f}, 
\label{eq:br} 
\end{equation} 
where the kinematic recoil factor $(1+\omega_{\pi^-}(q)/E_A(q))$ was 
approximated by its value for $q=q_0=101$ MeV/c and $A=11$, 
${\cal P}^{(\rm s)}_{i\to f}={\cal S}^{(\rm s)}_{i\to f}|F_{\rm DW}^{(\rm s)}
(q)|^2$, and ${\cal S}^{(\rm s)}_{i \to f}$ is given by Eq.~(\ref{eq:calPjj}). 
The summed rate, from an appropriately averaged hypernuclear ground-state 
doublet to all final states in the $1p$ shell through the $\pi^-$ decay 
$s$-wave amplitude, is given by 
\begin{equation} 
\frac{\sum_{f}\Gamma_{\pi^-}({\bar i} \to f)}{\Gamma_{\Lambda}}\approx 
0.743~\frac{\bar q}{q_0}~(1-\frac{1}{12}N_c^{(1p)})~
3|F_{\rm DW}^{(\rm s)}({\bar q})|^2, 
\label{eq:sumbr} 
\end{equation} 
where $|F_{\rm DW}^{(\rm s)}({\bar q})|^2$ is given by 
Eq.~(\ref{eq:DWFFinal}). We note that for hypernuclei based on $J_c=0$ 
nuclear cores, such as $_{\Lambda}^5{\rm He}$, $_{\Lambda}^9{\rm Be}$ and 
$_{~\Lambda}^{13}{\rm C}$, Eq.~(\ref{eq:sumbr}) gives the summed decay 
rate for the uniquely assigned $J_i=1/2$ hypernuclear ground state. 

As a brief aside we demonstrate the application of Eq.~(\ref{eq:sumbr}) 
to $_{\Lambda}^5{\rm He}$, the $\pi^-$ weak-decay rate of which was also 
determined by FINUDA \cite{Botta09}. Using the form-factor expression 
(\ref{eq:DWFFinal}) with ${\bar q} = 99.9$ MeV/c \cite{Gajewski69,Motoba91}, 
Eq.~(\ref{eq:sumbr}) gives a $1p$ contribution of 0.282 to the total decay 
rate, to which we add a $(2s-1d)$ contribution of 0.023 estimated from the 
calculation in Ref.~\cite{Motoba94} for $_{\Lambda}^7{\rm Li}$. Altogether, 
our estimated $\pi^-$ decay rate is 
$\frac{\Gamma_{\pi^-}^{\rm tot}({_{\Lambda}^5{\rm He}})}{\Gamma_{\Lambda}}=
0.305$, 
compared to $0.306\pm 0.060^{+0.025}
_{-0.020}$ \cite{Botta09} and to the KEK recent value $0.340 \pm 0.016$ 
\cite{Kameoka05}. 

\section{Results and Discussion} 

Using expressions (\ref{eq:br}) and (\ref{eq:sumbr}), our final results are 
listed in Table~\ref{tab:res} where we grouped the calculated $\pi^-$ decay 
rates into two partial sums: to the ground state and to nearby levels 
which share its underlying symmetry in the final nucleus, 
${\Gamma_{\pi^-}({_{\Lambda}^A}Z \to\overline{\rm g.s.})}/{\Gamma_{\Lambda}}$, 
and to excited levels that provide substantial decay width, 
${\Gamma_{\pi^-}({_{\Lambda}^A}Z \to {\rm exc.})}/{\Gamma_{\Lambda}}$. 
For the {\it total} rate ${\Gamma_{\pi^-}^{\rm tot}({_{\Lambda}^A}Z)}/
{\Gamma_{\Lambda}}$ we added where necessary the residual 
$\Lambda_{1s}\to p_{1p}$ contributions, as well as the small $(2s,1d)$ 
contribution of order $0.023 \pm 0.006$ from Ref.~\cite{Motoba94}. The choice 
of $\Lambda$ hypernuclear g.s. spins is discussed below for each hypernucleus 
separately. Unless stated differently, the nuclear wavefunctions are due to CK 
\cite{CK67} and the $\Lambda$ hypernuclear wavefunctions are considered in the 
weak coupling limit which provides a very reasonable approximation for summed 
rates, with estimated uncertainty of few percent. An educated guess for the 
overall theoretical uncertainty of our results is $10-15\%$.  

\begin{table} 
\begin{center} 
\caption{Calculated $\pi^-$ weak decay rates in $p$-shell hypernuclei 
where recent measurements by FINUDA are available {\protect{\cite{Botta09}}}. 
See text for discussion of the specified ground-state spins and for the 
approximations made. A comparison with the comprehensive calculations by 
Motoba et al. {\protect{\cite{Motoba94}}} is also provided} 
\label{tab:res} 
\begin{tabular}{lcccc}
\hline 
& ~$_{\Lambda}^7{\rm Li}({\frac{1}{2}}^+)$~ & ~$_{\Lambda}^9{\rm Be}
({\frac{1}{2}}^+)$~ & ~$_{~\Lambda}^{11}{\rm B}({\frac{5}{2}}^+)$~ & 
~$_{~\Lambda}^{15}{\rm N}({\frac{3}{2}}^+)$~ \\ 
\hline 
$q_{\rm g.s.}$ (MeV/c) & 108.1 & 97.0 & 106.1 & 98.4 \\ 
$q_{\rm exc.}$ (MeV/c) & 89.6 & 92.2 & 95.0 & -- \\ 
$\frac{\Gamma_{\pi^-}({_{\Lambda}^A}Z \to \overline{\rm g.s.})}
{\Gamma_{\Lambda}}$ &  0.261 & 0.102 & 0.079 & 0.064 \\ 
$\frac{\Gamma_{\pi^-}({_{\Lambda}^A}Z \to {\rm exc.})}{\Gamma_{\Lambda}}$ 
& 0.071 & 0.045 & 0.078 & -- \\ 
$\frac{\Gamma_{\pi^-}({_{\Lambda}^A}Z \to 2s,1d)}{\Gamma_{\Lambda}}$ 
& 0.024 & 0.029 & 0.018 & 0.022 \\ 
$\frac{\Gamma_{\pi^-}^{\rm tot}({_{\Lambda}^A}Z)}{\Gamma_{\Lambda}}$   
& 0.356 & 0.186 & 0.196 & 0.088 \\ 
$\frac{\Gamma_{\pi^-}^{\rm tot}({_{\Lambda}^A}Z)}
{\Gamma_{\Lambda}}$ \cite{Motoba94} & 0.304 & 0.172 & 0.213 & 0.090 \\ 
$\frac{\Gamma_{\pi^-}^{\rm tot}({_{\Lambda}^A}Z)}{\Gamma_{\Lambda}}$ 
exp. \cite{Botta09} & $0.353$ & $0.178$ & $0.249$ & $0.108$ \\  
$\pm$error &~$\pm 0.059^{+0.017}_{-0.013}$~&~$\pm 0.050^{+0.013}_{-0.008}$~& 
~$\pm 0.051^{+0.051}_{-0.023}$~&~$\pm 0.038^{+0.014}_{-0.013}$~\\
\hline 
\end{tabular} 
\end{center} 
\end{table} 

\subsection{$_{\Lambda}^7{\rm Li}$} 

Assuming $J^{\pi}(_{\Lambda}^7{\rm Li}_{\rm g.s.})={\frac{1}{2}}^+$ 
\cite{Sasao04} we find that the $_{\Lambda}^7{\rm Li}\to\pi^-~{^7{\rm Be}}$ 
weak decay is dominated by transitions to the lowest levels $^7{\rm Be}_
{\rm g.s.}({\frac{3}{2}}^-)$ and $^7{\rm Be}({\frac{1}{2}}^-;0.43~{\rm MeV})$, 
in agreement with Ref.~\cite{Motoba88}, with a summed strength 
${\cal S}^{(\rm s)}({_{\Lambda}^7{\rm Li}}({\frac{1}{2}}^+)\to {^7{\rm Be}}
(\overline{\rm g.s.}:0~\&~0.43~{\rm MeV})) = 2.32$. In the $LS$ coupling 
limit, which provides a good approximation in the beginning of the $1p$ shell, 
this summed strength equals ${\cal S}^{(\rm s)}(^2S_{\frac{1}{2}}\to 
{^2P_{J_f}})= 5/2$ and is distributed according to $(2J_f+1)$ whereas for 
$J^{\pi}(_{\Lambda}^7{\rm Li}_{\rm g.s.})={\frac{3}{2}}^+$, owing to the 
spin-nonflip nature of the dominant $s$-wave $\pi$ decay, the $LS$ limit 
gives ${\cal S}^{(\rm s)}(^4S_{\frac{3}{2}}\to {^2P_{J_f}})=0$. The 
prominence of the ${^7{\rm Be}}(0~\&~0.43~{\rm MeV})$ final states in the 
$_{\Lambda}^7{\rm Li} \to \pi^- ~ {^7{\rm Be}}$ weak decay FINUDA spectrum 
in Ref.~\cite{Botta09} clearly rules out assigning 
$J^{\pi}(_{\Lambda}^7{\rm Li}_{\rm g.s.})={\frac{3}{2}}^+$. 
Additional strength for $J^{\pi}(_{\Lambda}^7{\rm Li}_{\rm g.s.}) = 
{\frac{1}{2}}^+$, with ${\cal S}^{(\rm s)}=0.98$ using CK wavefunctions, 
goes to $^7{\rm Be}$ levels around 11 MeV (${\cal S}^{(\rm s)}= 1$ in the 
$LS$ limit). Altogether, the total strength to $p$-shell nuclear states is 
${\cal S}^{(\rm s)}_{\rm tot}=3.29$ (7/2 in the $LS$ limit). The calculated 
decay rates listed in the table agree very well with the shape and the 
strength of the measured FINUDA spectrum, confirming the spin assignment 
$J^{\pi}(_{\Lambda}^7{\rm Li}_{\rm g.s.})={\frac{1}{2}}^+$ \cite{Sasao04}. 
Our calculated total rate is $17\%$ higher than previously calculated 
\cite{Motoba94}. 

\subsection{$_{\Lambda}^9{\rm Be}$} 

For this hypernucleus, with nuclear core spin $J_c^{\pi}=0^+$, we use the 
sum-rule expression (\ref{eq:sumrule}) to obtain a transition strength 
${\cal S}^{(\rm s)}_{\rm tot}=2$. The leading transitions are to 
$^9{\rm B}_{\rm g.s.}({\frac{3}{2}}^-)$ and to the ${\frac{1}{2}}^-$ excited 
state at 2.75 MeV, both of which saturate the sum-rule strength in the $LS$ 
limit. Our calculated total rate agrees well with the FINUDA measurement 
\cite{Botta09} within the experimental uncertainty, and is $8\%$ higher 
than in previous calculations \cite{Motoba94}. 

\subsection{$_{~\Lambda}^{11}{\rm B}$} 

Assuming $J^{\pi}(_{\Lambda}^{11}{\rm B}_{\rm g.s.})={\frac{5}{2}}^+$ 
\cite{ZD75}, the calculated summed strength is 
${\cal S}^{(\rm s)}_{\rm tot}=2.04$, with the largest contributions due to 
$^{11}{\rm C}_{\rm g.s.}({\frac{3}{2}}^-)$ (${\cal S}^{(\rm s)}=0.73$) and to 
the ${\frac{7}{2}}^-$ excited state at 6.48 MeV (${\cal S}^{(\rm s)}=0.94$). 
The calculated total decay rate given in the table is lower by $8\%$ than 
that calculated previously \cite{Motoba94}, and by just $1\sigma_{\rm stat}$ 
than the measured mean value. Given an estimated $10\%$ error on our 
calculated rate, there is certainly no disagreement between the present 
calculation and experiment. 

If the spin of $_{\Lambda}^{11}{\rm B}$ were ${\frac{7}{2}}^+$, there 
would have been no $s$-wave transition to $^{11}{\rm C}_{\rm g.s.}$, 
the transition to the ${\frac{7}{2}}^-$ excited state at 6.48 MeV 
would have been $5-6$ times weaker than it is supposed to be for 
$J^{\pi}(_{\Lambda}^{11}{\rm B}_{\rm g.s.})={\frac{5}{2}}^+$, and 
the dominant transition would have been to the ${\frac{5}{2}}^-$ 
excited state at 8.42 MeV. A comparison with the shape of the measured 
spectrum rules out this possibility. Furthermore, assigning to all the 
$\Lambda_{1s}\to p_{1p}$ transitions a fixed value $q_{\rm exc.}=91.6$ 
MeV/c appropriate to this dominant transition to the ${\frac{5}{2}}^-$ 
level, and adding the $\Lambda_{1s}\to p_{2s,1d}$ strength, one obtains 
${\Gamma_{\pi^-}^{\rm tot}(_{\Lambda}^{11}{\rm B}({\frac{7}{2}}^+))}
/{\Gamma_{\Lambda}}=0.101$, about half that calculated for spin 
${\frac{5}{2}}^+$ as listed in Table~\ref{tab:res}, 
again in clear disagreement with the data. This provides a firm confirmation 
of the assignment $J^{\pi}(_{\Lambda}^{11}{\rm B}_{\rm g.s.})={\frac{5}{2}}^+$ 
made \cite{ZD75} on the basis of $\pi^-$ decays observed in emulsion, not all 
of which  uniquely associated with $_{\Lambda}^{11}{\rm B}$ \cite{Juric73}. 
Although this assignment is in good accord with previously reported total 
$\pi^-$ decay rates \cite{Sato05} and with several theoretical interpretations 
of other data (e.g. Refs.~\cite{DDT86,MGDD85,Mill08}), the present FINUDA 
measurement of the complete decay spectrum provides the best experimental 
evidence for $J^{\pi}(_{\Lambda}^{11}{\rm B}_{\rm g.s.})={\frac{5}{2}}^+$.  

\subsection{$_{~\Lambda}^{15}{\rm N}$} 

The ground-state spin of $_{~\Lambda}^{15}{\rm N}$ has not been determined 
experimentally. The first realistic discussion of spin dependence in 
$p$-shell $\Lambda$ hypernuclei placed the ${\frac{1}{2}}^+$ g.s. doublet 
member barely above the ${\frac{3}{2}}^+$ ground state, at 14 keV excitation, 
so both of these levels are practically degenerate and would decay weakly 
\cite{MGDD85}. This near degeneracy reflects the competition between the 
$\Lambda N$ spin-spin interaction (favoring $J^{\pi}={\frac{3}{2}}^+$) and 
the $\Lambda N$ tensor interaction (favoring $J^{\pi}={\frac{1}{2}}^+$) in 
the $p_{\frac{1}{2}}$ subshell. The most recent theoretical update 
\cite{Mill08} places the ${\frac{1}{2}}^+$ excited level about 90 keV above 
the ${\frac{3}{2}}^+$ ground state. The spin ordering cannot be determined 
from the $\gamma$-ray spectrum of $_{~\Lambda}^{15}{\rm N}$ measured in the 
BNL-E930 experiment \cite{Ukai08}. Our calculated $\pi^-$ weak-decay rates 
to final states in $^{15}$O confirm the conclusion of Ref.~\cite{Motoba88} 
that $^{15}{\rm O_{g.s.}}({\frac{1}{2}}^-)$ nearly saturates the total 
$\Lambda_{1s}\to p_{1p}$ spectroscopic strength. This near saturation follows 
from the calculated decay rates for $_{~\Lambda}^{15}{\rm N}\to \pi^-~{^{15}
{\rm O_{g.s.}}}({\frac{1}{2}}^-)$ listed in Table~\ref{tab:L15N}. Indeed, the 
saturation is complete in the jj coupling limit, where ${\cal S}^{(\rm{s,p})}_
{jj}(\overline{\rm g.s.}\to {^{15}{\rm O_{g.s.}}}({\frac{1}{2}}^-))=1/2$ for 
the $(2J_i+1)$-average of strengths ${\cal S}^{(\rm {s,p})}_{jj}$, agreeing 
with the sum-rule value $\sum_{f}{\cal S}^{(\rm {s,p})}_{{\bar i}\to f} = 
3(1-\frac{1}{12}N_c^{(1p)})$ given in Eqs.~(\ref{eq:sumrule}) and 
(\ref{eq:sumrulepwave}) for $s$-wave and $p$-wave amplitudes, respectively. 
The saturation holds approximately in the more realistic CK model, for which 
the strengths listed in the table give ${\cal S}^{(\rm {s,p})}_{CK}(\overline
{\rm g.s.}\to {^{15}{\rm O_{g.s.}}}({\frac{1}{2}}^-))=0.486$ for both $s$-wave 
and $p$-wave amplitudes. Very little spectroscopic strength is therefore left 
to the $p_{\frac{3}{2}}^{-1}p_{\frac{1}{2}}$ excited state 
$^{15}{\rm O}({\frac{1}{2}}^-;6.18~{\rm MeV})$ 
which may safely be neglected. We note that the $s$-wave strength for 
$J^{\pi}(_{\Lambda}^{15}{\rm N}_{\rm g.s.})={\frac{1}{2}}^+$ 
is four times weaker in the jj coupling limit than for spin ${\frac{3}{2}}^+$. 
Using realistic wavefunctions leads to even greater suppression, by a factor 
about 14 for CK and by a similar factor for Millener's \cite{Mill08} recent 
re-analysis of nuclei in the $p$-shell.{\footnote{Millener's $^{14}{\rm N}
_{\rm g.s.}$ wavefunction is dominated to $93\%$ by a $^{3}D_1$ component 
which for $J^{\pi}(_{\Lambda}^{15}{\rm N})={\frac{1}{2}}^+$ leads in the weak 
coupling limit to a single $LS$ hypernuclear component $^{4}D_{\frac{1}{2}}$. 
Since the transition $_{~\Lambda}^{15}{\rm N}({^{4}D_{\frac{1}{2}}}) \to 
{^{15}{\rm N}}({^{2}P_{\frac{1}{2}}})$ requires spin-flip, it is forbidden 
for the $\pi^-$-decay dominant $s$-wave amplitude.}} The suppression of 
$_{~\Lambda}^{15}{\rm N}({\frac{1}{2}}^+) \to \pi^-~{^{15}{\rm O_{g.s.}}}$ 
with respect to 
$_{~\Lambda}^{15}{\rm N}({\frac{3}{2}}^+) \to \pi^-~{^{15}{\rm O_{g.s.}}}$ 
was overlooked in Ref.~\cite{Motoba88}. 

Finally, since a single transition 
$_{~\Lambda}^{15}{\rm N} \to \pi^-~{^{15}{\rm O_{g.s.}}}$ dominates the 
$s$-wave $\pi^-$ decay, irrespective of which value holds for 
$J^{\pi}(_{\Lambda}^{15}{\rm N}_{\rm g.s.})$, it is necessary 
to check explicitly the effect of the weaker $p$-wave $\pi^-$ decay
which might beef up the suppressed $s$-wave strength for 
$J^{\pi}(_{\Lambda}^{15}{\rm N}_{\rm g.s.})={\frac{1}{2}}^+$. 
This is indeed the case, as shown in Table~\ref{tab:L15N}, where the 
combined $s$- plus $p$-wave decay rate for 
$J^{\pi}(_{\Lambda}^{15}{\rm N}_{\rm g.s.})={\frac{1}{2}}^+$ is now 
only a factor of 2 or 3 (for jj and CK, respectively) weaker than for 
$J^{\pi}(_{\Lambda}^{15}{\rm N}_{\rm g.s.})={\frac{3}{2}}^+$, no longer 
by a factor of over 10 as for $s$-wave decay alone. 
Including the additional 
very small partial rate due to $^{15}{\rm O}({\frac{3}{2}}^-)$ and the 
contribution due to ($2s-1d$) final states, our {\it total} 
$_{~\Lambda}^{15}{\rm N} \to \pi^-~{^{15}{\rm O}}$ decay rate assumes 
the values 
\begin{equation} 
\frac{\Gamma_{\pi^-}^{\rm tot}(_{\Lambda}^{15}{\rm N}({\frac{3}{2}}^+))}
{\Gamma_{\Lambda}}=0.080,~~~~~
\frac{\Gamma_{\pi^-}^{\rm tot}(_{\Lambda}^{15}{\rm N}({\frac{1}{2}}^+))}
{\Gamma_{\Lambda}}=0.040,  
\label{eq:BR15} 
\end{equation}
depending on the hypernuclear ground-state spin. It is worth noting how 
little the introduction of the $\pi^-$ decay $p$-wave amplitude affected the 
total rate for $J^{\pi}(_{\Lambda}^{15}{\rm N}_{\rm g.s.})={\frac{3}{2}}^+$, 
from a rate 0.088 in Table~\ref{tab:res} without explicit consideration of 
$p$-waves, to 0.080 upon explicitly including $p$-waves. This latter value 
is $11\%$ lower than the decay rate 0.090 calculated by Motoba et 
al.~\cite{Motoba94} and is short of the FINUDA measured total rate by 
just $1\sigma_{\rm stat}$. In contrast, the calculated decay rate for 
$J^{\pi}(_{\Lambda}^{15}{\rm N}_{\rm g.s.})={\frac{1}{2}}^+$ 
in Eq.~(\ref{eq:BR15}) is $2\sigma_{\rm stat}$ below the measured 
mean value. In this respect we disagree with the conclusion deduced from 
Ref.~\cite{Motoba88} that the total $\pi^-$ decay rate of 
$_{~\Lambda}^{15}{\rm N}$ depends weakly on the value of its ground-state 
spin.{\footnote{The calculation of Ref.~\cite{Motoba94} yields 
$\frac{\Gamma_{\pi^-}^{\rm tot}(_{\Lambda}^{15}{\rm N}({\frac{1}{2}}^+))}
{\Gamma_{\Lambda}}=0.074$ compared with 0.090 for ground-state spin-parity 
${\frac{3}{2}}^+$ (T. Motoba, private communication).}} A similar disagreement 
holds with respect to the two values of calculated partial rates listed in 
Table~\ref{tab:L15N} for $_{~\Lambda}^{15}{\rm N}({\frac{1}{2}}^+)\to \pi^-~
{^{15}{\rm O_{g.s.}}}({\frac{1}{2}}^-)$, where our calculated value is by 
over $2\sigma$ lower than the measured partial rate. We consider our present 
calculation, in conjunction with the FINUDA measurement \cite{Botta09}, 
a determination of the ground-state spin of $_{~\Lambda}^{15}{\rm N}$, 
namely $J^{\pi}(_{\Lambda}^{15}{\rm N}_{\rm g.s.})={\frac{3}{2}}^+$.  

\begin{table} 
\begin{center} 
\caption{$_{~\Lambda}^{15}{\rm N}(J^{\pi}) \to \pi^-~{^{15}{\rm O_{g.s.}}}
({\frac{1}{2}}^-)$ calculated $s$-wave and $p$-wave spectroscopic factors 
${\cal S}$, and summed $s$-wave plus $p$-wave weak decay rates, compared with 
FINUDA's measured decay rates \cite{Botta09} and with those calculated by 
Motoba et al.~{\cite{Motoba94}} (T.~Motoba, private communication.)} 
\label{tab:L15N} 
\begin{tabular}{lcc|ccc|ccc} 
\hline 
$_{~\Lambda}^{15}{\rm N}$ & exp. \cite{Botta09} & calc. \cite{Motoba94} & 
\multicolumn{3}{c}{jj} & \multicolumn{3}{c}{CK} \\ 
$J^{\pi}$ & $\frac{\Gamma_{\pi^-}}{\Gamma_{\Lambda}}$ & 
$\frac{\Gamma_{\pi^-}}{\Gamma_{\Lambda}}$ & ${\cal S}^{(\rm s)}$ 
& ${\cal S}^{(\rm p)}$ & $\frac{\Gamma_{\pi^-}}{\Gamma_{\Lambda}}$ & 
${\cal S}^{(\rm s)}$ & ${\cal S}^{(\rm p)}$ & 
$\frac{\Gamma_{\pi^-}}{\Gamma_{\Lambda}}$ \\ \hline 
${\frac{3}{2}}^+$ & $0.072 \pm 0.024$ & 0.065 & $\frac{2}{3}$ & 0 & 0.053 \ & 
\ 0.703 \ & \ 0.016 \ & \ 0.057 \\  
${\frac{1}{2}}^+$ & $0.072 \pm 0.024$ & 0.050 & $\frac{1}{6}$ & $\frac{3}{2}$ 
& 0.029 \ & \ 0.052 \ & \ 1.426 \ & \ 0.019 \\  
\hline 
\end{tabular} 
\end{center} 
\end{table}

\section{Sum-rule applications} 

The sum rule Eq.~(\ref{eq:sumbr}) for the $s$-wave amplitude and a similar 
one for the $p$-wave amplitude involve a $(2J_i+1)$-average over the initial 
hypernuclear ground-state configuration, and as such cannot be applied to any 
of the total mesonic weak decay rates measured by FINUDA, which relate to 
specific g.s. values $J_i$, except for the $J_c=0$ core $_{\Lambda}^9{\rm Be}$ 
hypernucleus. However, it may be applied to available theoretical evaluations 
in order to provide consistency checks. This assumes that the $A$ dependence 
of the closure ${\bar q}$ value is weak, typically $10\%$ or less, and also 
that the DW integrals depend weakly on $A$ apart from the dependence on 
${\bar q}$ in Eq.~(\ref{eq:DWFFinal}). Sum-rule values are shown in 
Table~\ref{tab:sumrule} for the present $s$-wave calculation as well as for 
the comprehensive calculations by Motoba et al. \cite{Motoba94}, where we used 
the $(1-N_c^{(1p)}/12)$ $A$-dependence from Eq.~(\ref{eq:sumbr}) to normalize 
the calculated $(2J_i+1)$-averaged $\Lambda_{1s} \to p_{1p}$ hypernuclear 
$\pi^-$ weak decay rates to that of $_{~\Lambda}^{15}{\rm N}$ 
($N_c^{(1p)}=10$), in units of the calculated $(2J_i+1)$-averaged 
$\Lambda_{1s} \to p_{1p}$ $_{~\Lambda}^{15}{\rm N}$ $\pi^-$ weak decay rate. 
For completeness, we added $_{~\Lambda}^{13}{\rm C}$ so as to present a full 
range of odd-$A$ hypernuclei in the $1p$ shell. 

\begin{table} 
\begin{center} 
\caption{Calculated $(2J_i+1)$-averaged $\Lambda_{1s}\to p_{1p}$ hypernuclear 
$\pi^-$ weak decay rates divided by $(6-N_c^{(1p)}/2)$ and by the calculated 
$(2J_i+1)$-averaged $\Lambda_{1s}\to p_{1p}$ $_{~\Lambda}^{15}{\rm N}$ $\pi^-$ 
weak decay rate, see text. Given in the column before last are the mean and 
standard deviation of the averaged rates for the four preceding hypernuclei}  
\label{tab:sumrule} 
\begin{tabular}{lcccccc} 
\hline 
Ref. & $_{\Lambda}^7{\rm Li}$ & $_{\Lambda}^9{\rm Be}$ & 
$_{~\Lambda}^{11}{\rm B}$ & $_{~\Lambda}^{13}{\rm C}$ & 
\ mean ($_{\Lambda}^7{\rm Li} - {_{~\Lambda}^{13}{\rm C}}$) \ & 
$_{~\Lambda}^{15}{\rm N}$ \\ 
\hline 
\cite{Motoba94} & \ 0.635 \ & \ 0.577 \ & \ 0.748 \ & \ 0.633 \ & 
\ $0.648 \pm 0.072$ \ & \ 1 \\ 
present & \ 0.945 \ & \ 0.873 \ & \ 0.920 \ & \ 0.813 \ & 
\ $0.888 \pm 0.058$ \ & \ 1 \\  
\hline 
\end{tabular} 
\end{center} 
\end{table} 

Table~\ref{tab:sumrule} demonstrates that the sum-rule-normalized 
$\Lambda_{1s}\to p_{1p}$ hyprnuclear $\pi^-$ weak decay rates of 
$_{\Lambda}^7{\rm Li}$, $_{\Lambda}^9{\rm Be}$, $_{~\Lambda}^{11}{\rm B}$ 
and $_{~\Lambda}^{13}{\rm C}$ are represented fairly well by a mean value, 
with only one of the species ($_{~\Lambda}^{11}{\rm B}$ for the calculation 
of Ref.~\cite{Motoba94} and $_{~\Lambda}^{13}{\rm C}$ in the present one) 
departing by somewhat over one standard deviation. 
Percentage-wise, it represents $8\%$ in our calculation and $15\%$ for 
Ref.~\cite{Motoba94}. However, the value 1 for $_{~\Lambda}^{15}{\rm N}$ 
deviates from the corresponding mean by five standard deviations in the 
calculation of Ref.~\cite{Motoba94}, whereas it deviates by `merely' two 
standard deviations in our calculation. This provides a circumstantial 
argument in favor of the correctness of the present $_{~\Lambda}^{15}{\rm N}$ 
calculation in which even the seemingly substantial $2\sigma$ deviation for 
$_{~\Lambda}^{15}{\rm N}$ amounts only to $13\%$ departure from the mean rate. 
In contrast, the $5\sigma$ deviation for $_{~\Lambda}^{15}{\rm N}$ in the 
calculation of Ref.~\cite{Motoba94} amounts to a huge $54\%$ departure from 
the mean which is hard to accept. 

\section{Summary} 

We have checked and updated some of the pioneering calculations of mesonic 
weak decay of hypernuclei by Motoba et al. \cite{Motoba88,Motoba92,Motoba94}. 
Our added value is in providing an explicit spectroscopic expression 
Eq.~(\ref{eq:calPjj}) and a related sum rule Eq.~(\ref{eq:sumrule}) for 
$\Lambda_{1s}\to p_{1p}$ $s$-wave transitions. The neglect of $p$-wave 
transitions for summed spectra is justified to better than $10\%$. 
By comparing the calculated rates with the FINUDA measured rates 
\cite{Botta09} we have pointed out that these recent data 
confirm the hypernuclear ground-state spin-parity assignments 
$J^{\pi}(_{\Lambda}^{7}{\rm Li_{g.s.}})={\frac{1}{2}}^+$ and 
$J^{\pi}(_{~\Lambda}^{11}{\rm B_{g.s.}})={\frac{5}{2}}^+$, while providing 
a first experimental determination for $_{~\Lambda}^{15}{\rm N}$: 
$J^{\pi}(_{~\Lambda}^{15}{\rm N_{g.s.}})={\frac{3}{2}}^+$. 
     
Whereas good agreement between calculations and experiment for 
$_{\Lambda}^7{\rm Li}$ and $_{\Lambda}^9{\rm Be}$ has been demonstrated, 
the total $\pi^-$ decay rates measured by FINUDA for $_{~\Lambda}^{11}{\rm B}$ 
and $_{~\Lambda}^{15}{\rm N}$ are higher than our calculated total rates by 
about $1\sigma_{\rm stat}$, which might suggest the presence of other 
hypernuclear weak decays than the ones studied here. We note that in both of 
these cases the formation of $_{\Lambda}^AZ$ proceeds by proton emission from 
$_{\Lambda}^{A+1}(Z+1)$ continuum where contamination by nearby thresholds 
($n,\alpha,\cdots$) could introduce a bias. Such a possibility has been 
analyzed for $_{~\Lambda}^{15}{\rm N}$ in detail by FINUDA \cite{Botta09}, 
but without providing (yet) a quantitative measure of the additional 
hypernuclear components that are likely to contribute to the spectral shape 
of the measured weak decay. All in all, we have demonstrated a satisfactory 
level of agreement between calculations of mesonic weak-decay rates of 
$p$-shell $\Lambda$ hypernuclei and the recent FINUDA measurements.

\section*{Acknowledgements} 
I would like to thank Elena Botta, Tullio Breessani, and Stefania Bufalino 
who introduced me to the recent measurements of hypernuclear weak decays by 
the FINUDA Collaboration, to John Millener for providing his nuclear and 
hypernuclear wavefunctions for $A=14,15$, and to Toshio Motoba for providing 
details of the calculations reported in Ref.~\cite{Motoba94}.

\end{document}